\documentclass[aps,twocolumn,prl,tightenlines,floatfix]{revtex4}
\usepackage[dvips]{graphicx}
\usepackage[english]{babel}
\usepackage{amsmath}
\usepackage{amssymb}

\newcommand{\bs}{\begin{split}}
\newcommand{\es}{\end{split}}
\newcommand{\be}{\begin{equation}}
\newcommand{\ee}{\end{equation}}
\newcommand{\ba}{\begin{eqnarray}}
\newcommand{\ea}{\end{eqnarray}}

\def\ket#1{|#1\rangle}

\begin{document}

\title{Signatures of superfluidity for Feshbach-resonant Fermi gases}
\author{J. Kinnunen, M. Rodr\'{\i}guez, and P. T\"orm\"a}
\affiliation{Department of Physics, NanoScience Center, P.O.Box 35, FIN-40014 University of Jyv\"askyl\"a, Finland}

\begin{abstract}
We consider atomic Fermi gases where Feshbach resonances can be used to continuously tune the system from weak to strong interaction regime, allowing to scan the whole BCS-BEC crossover. We show how a probing field transferring atoms out of the superfluid can be used to detect the onset of the superfluid transition in the high-$T_c$ and BCS regimes. The number of transferred atoms, as a function of the energy given by the probing field, peaks at the gap energy. The shape of the peak is asymmetric due to the single particle excitation gap. Since the excitation gap includes also a pseudogap contribution, the asymmetry alone is not a signature of superfluidity. Incoherent nature of the non-condensed pairs leads to broadening of the peak. The pseudogap and therefore the broadening decay below the critical temperature, causing a drastic increase in the asymmetry. This provides a signature of the transition. 
\end{abstract}

\maketitle

Novel type of superfluid, a Bose-Einstein condensate of molecules
consisting of fermionic alkali atoms, has been created recently
\cite{expe,expethe}.
Molecules are formed by a Feshbach resonance in the range of parameters where the interaction between the atoms is repulsive. 
By tuning magnetic fields the resonance can be crossed and attractive
interactions realized. Crossing of the resonance can be done adiabatically, in temperatures
lower than the critical temperature
\cite{expenew}, and fermion pairing investigated \cite{expenewnew}. This opens up unprecedented prospects for studying the BCS-BEC crossover from a superfluid of momentum-space pairs to a BEC of real-space dimers. The experiments \cite{expe,expenew} demonstrate the existence of a superfluid in the BEC limit of the crossover. In
this letter we propose a method to observe an explicit signature of
the superfluid transition also in the intermediate (high-$T_c$) and BCS regimes. 

Fermionic alkali atoms in the vicinity of a Feshbach resonance are predicted to form superfluids with three components: 1) fermionic Cooper-pairs, 2) molecular bosons formed out of the fermions, 3) non-condensed fermionic pairs corresponding to a pseudogap. Non-condensed pairs and molecules exist also above the critical temperature. The system can be described by the resonance superfluidity theory \cite{ressup,Kokkelmans2002}. In the BEC limit, the superfluid transition is evident like for an atomic BEC, for instance via the bimodal density distribution \cite{expe,expenew}. In the high-$T_c$ and BCS regimes such a strong effect on the density is not expected. Moreover, strong pseudogap or mean field may be almost indistinguishable from the effects of superfluidity. Several signatures of superfluidity for fermionic atoms have been proposed in \cite{paivi,signatures}, however, these works assume simple BCS scenario and do not take into account the pseudogap and the bosonic molecules. 
Here we discuss a method that can give a clear signature of the resonance superfluidity transition. It is experimentally feasible since a similar technique has already been used to study mean-field effects \cite{rf}. The method is based on transferring atoms from the superfluid state into a normal state using electromagnetic fields. There is an analogy to imposing current over a superconductor-normal metal interface. Instead of a localized tunneling barrier, the transfer is
caused by a homogeneous field, enforcing exact momentum
conservation. This leads to a resonance-type of response as a function of the applied energy, in
contrast to superconductors where the current grows steadily as a function of
voltage once the gap energy is exceeded \cite{Mahan}. 

Interacting gas of two fermionic components can be formed by trapping atoms in two different hyperfine states, let us label the states
by $\ket{g}$, $\ket{g'}$. The strength of the interaction between atoms in
$\ket{g}$ and $\ket{g'}$ is tuned by
magnetic (or optical) fields using a Feshbach resonance. The atoms provide also other hyperfine
states (label $\ket{e}$) for which the interatomic scattering lengths
can be very different from that between $\ket{g}$ and $\ket{g'}$. This is
utilized in \cite{rf} to observe mean field effects: atoms are resonantly transferred, say, between $\ket{g}$ and $\ket{e}$, and the
difference in the mean
fields experienced by $\ket{g}$ and $\ket{e}$
causes an energy shift in the position of
the resonance. Similarly, for a superfluid formed of atoms in the $\ket{g}$ and $\ket{g'}$ states, the BCS-gap in the single particle
excitation spectrum is predicted to cause a shift in the
position of the resonance \cite{paivi}. The main qualitative difference to the mean-field shift is the asymmetric shape of the peak in the superfluid case. Here we show that a pseudogap can lead to a shifted, asymmetric peak as well. Therefore asymmetry alone is not a signature of superfluidity. The transition, however, is signaled by an abrupt change in the amount of the asymmetry.  

The system is described by the Hamiltonian
\begin{eqnarray}
\label{eq:hamil}
&H&= \sum_{k,\sigma} \varepsilon^{\sigma}_{k} c^{\sigma\dagger}_{k}
c^{\sigma}_{k} + \sum_{q}(E_q^{0}+\nu)b_{q}^\dagger b_{q} \nonumber\\
&&+\sum_{q,k,k'}U(k,k')c_{q/2+k}^{g\dagger}c_{q/2-k}^{g'\dagger}
c_{q/2-k'}^{g'}c_{q/2+k'}^{g} \nonumber\\
&&+ \sum_{q,k}(g(k)b_q^\dagger c^g_{q/2-k}
c^{g'}_{q/2+k}+ h.c.)+ H_T , \nonumber
\end{eqnarray}
\begin{eqnarray}
&H_T&=\sum_k\frac{\delta}{2}\left(c_k^{e\dagger}c_k^e-
c_k^{ g\dagger}c_k^g  -  b_k^{\dagger}b_k \right)
\nonumber\\
 &&+\sum_{k,l}
\left( M_{kl} c^{g\dagger}_k c_l^e + h.c. \right)
+ \sum_{klq} \left( D_{qkl}
b_q^\dagger c^{g'}_k c_l^e+ h.c. \right)  \nonumber
\ea
The operators $c_k^\sigma$ correspond to fermionic atoms in various hyperfine states  ($\sigma=\{e,g,g'\}$) and $b_q$ to spinless bosonic molecules formed out of atoms in states $\ket{g}$ and $\ket{g'}$. The fermion and boson energies are
$\varepsilon_k=k^2/2m$ and $E_q^0=q^2/2M$, respectively, with $m$ the fermion mass, $M=2m$ and $\hbar\equiv 1$. A momentum cutoff is used: 
$U(k,k')=U\varphi_k\varphi_{k'}$, $g(k)=g\varphi_k$, where $\varphi_k^2=\exp{-(k/K_c)^2}$. The effect of the Feshbach resonance can be incorporated to a point-like interaction between the atoms, as in $H$, by using renormalized parameters $U$, $g$ and $\nu$ \cite{Kokkelmans2002}. The corresponding unrenormalized parameters are:  
$U_0=4\pi a_s/m$ is the fermion-fermion interaction, with $a_s$ the scattering length, $g_0$ is the boson-fermion coupling, and $\nu_0$ the magnetic field detuning from the Feshbach resonance. Positive $\nu_0$ corresponds to attractive interactions between the atoms, negative $\nu_0$ to repulsive. At the resonance, $\nu_0=0$, the system is in the high-$T_c$ region. The limits $\nu_0 \longrightarrow \infty$ and $\nu_0 \longrightarrow - \infty$ are the extreme BCS and BEC limits, respectively. 
The parameters used throughout the paper are $g_\mathrm{0} = -5.0\,E_\mathrm{F}$ and
$U_\mathrm{0} = 0.05\,E_\mathrm{F}$, corresponding to the density of atoms $n = 10^{13} \, {\mathrm{cm}}^{-3}$
and background scattering length $a_\mathrm{s} = 176\,a_\mathrm{0}$.
The initial numbers of atoms in states $\ket{g}$ and $\ket{g'}$ are equal, therefore the corresponding chemical potentials
$\mu_g=\mu_{g'}\equiv\mu$. The state $\ket{e}$ is assumed initially empty, that is, $\mu_e=0$.

The observable of interest is the current of atoms from the state $\ket{g}$
to state $\ket{e}$, $I=<\dot N_e>$. 
The effect of the electromagnetic field transferring atoms out of the superfluid, i.e. 
creating the superfluid-normal state interface, is described by $H_T$.
To match the energy difference between the considered hyperfine states, the field can be 
either an rf-field or laser field creating a Raman transition. In describing the 
interaction of the field with the molecule (formed of fermions in states $\ket{g}$ and 
$\ket{g'}$), we have considered only the interaction of the field with the two-level 
system $\ket{g}$ --- $\ket{e}$. This approximation is made having in mind large, 
extremely weakly bound molecules \cite{expe,expenew}. We refer to the detuning 
$\delta = (\omega_e - \omega_g) - \omega_{field}$ by probe field detuning, not to 
confuse it with the Feshbach detuning $\nu_0$ of the magnetic field.

The particle current $I = <\dot N_\mathrm{e}> = i <\left[ H, N_\mathrm{e}\right] >$ is calculated treating $H_\mathrm{T}$ as a perturbation and neglecting terms of higher than second order.
The solutions for the equilibrium parameters, chemical potential and the excitation gap, are needed for calculating the current. 
We apply the $t$-matrix approach to describe the pairing interaction,
adopting the procedure used in Ref.~\cite{Stajic2003} for solving the
relevant equations: gap, pseudogap and number equations.
The chemical potential $\mu$, pseudogap $\Delta_\mathrm{pg}$ and the superfluid order parameter $\tilde \Delta_\mathrm{sc}$ are obtained from the self-consistent solutions. 

Physically, one may interpret the pseudogap as fermion pairs which have a finite net momentum 
and are formed above the critical temperature (thus the name preformed pairs). At the critical 
temperature, these pairs condense to the zero net-momentum state. The origin of the pseudogap 
in the many-body theory are fluctuations of the finite momentum order parameter 
$\langle c_{-k}^g c_{k+q}^{g'}\rangle$ around the zero-momentum mean 
$\langle c_{-k}^g c_k^{g'} \rangle $. The zero-momentum mean is the usual BCS order parameter, 
and has a non-zero value below the transition temperature. For finite momenta $q$, the mean 
$\langle c_{-k}^g c_{k+q}^{g'}\rangle$ is always zero. However, it has a non-zero variance, 
corresponding to fluctuations, also above $T_c$. This leads to the pseudogap. The $t$-matrix
for the system can be written as a sum of a superfluid term $\tilde{\Delta}_{sc}^2$ containing 
a pole at $Q=0$, and a pseudogap contribution that includes the rest of the 
$t$-matrix, $\Delta_\mathrm{pg}^2 = -\sum_{Q \neq 0} t (Q)$~\cite{Chen1998}. 
The total excitation gap is given by $\Delta^2 = \tilde{\Delta}_{sc}^2 + \Delta_{pg}^2$.
Pseudogap has a considerable effect in the high-$T_\mathrm{c}$ regime but vanishes in 
the extreme BCS limit.      



The pseudogap and the order parameter are qualitatively different and in 
principle can be distinguished from each other through life-time effects. 
The incoherent nature of the propagating, non-condensed pairs affects 
scattering processes and thus incorporates a finite broadening in the self-energy $\Sigma_\mathrm{pg}$. This is taken into account by replacing the 
self-energy by the model self-energy~\cite{Maly1999b}
\ba
\label{eq:selfenergy2}
  \Sigma (K) = \frac{\tilde \Delta_\mathrm{sc}^2 \varphi_k^2}{i\omega + \epsilon_k} 
     + \frac{\Delta_\mathrm{pg}^2 \varphi_k^2}{i \omega+\epsilon_k+i \gamma}, 
\ea
where $\gamma$ is the finite broadening. Using the model self-energy, we have 
plotted the fermionic density of states 
$N(\omega) = -2 \sum_k \mathrm{Im} G(\omega+i0,k)$ in the inset of Fig.~\ref{fig:cur}
for several temperatures. We have chosen the Feshbach detuning $\nu_\mathrm{0} = 2.0\,E_\mathrm{F}$ (attractive interactions)
and the broadening 
$\gamma = \Delta (T_\mathrm{c}) \approx 0.6\,E_\mathrm{F}$.
The plot shows the widening of the peak at $\omega = -\Delta$ as the temperature
increases, caused mainly by the emergence of the pseudogap. At lower temperatures,
a peak appears also at $\omega = +\Delta$, and in the limit $T \rightarrow 0$
the density of states approaches the BCS limit of two peaks. 

The current $I = <\dot N_\mathrm{e}> = i <\left[ H, N_\mathrm{e}\right] >$ can now be calculated. Denoting
$A(t) = \sum_{kl} M_{kl} c_k^{\mathrm{g}\dagger} c_l^\mathrm{e}$ and 
$B(t) = \sum_{qkl} D_{qkl} b_q^\dagger c_l^\mathrm{g'} c_k^\mathrm{e}$,
where $M_{kl}$, $D_{qkl}$ are the coupling constants, the correlations of the form 
$<AA>$, $<BB>$, $<BA>$ and $<B^\dagger A>$ 
(and $h.c.$) are neglected. The first three contain correlations of the form $\langle c^e c^e \rangle$ which are identically zero for a normal state. The last one corresponds to a Josephson current between the molecular and fermion superfluids, mediated by the $\ket{e}$ state \cite{footnote1}. 

Writing the total current as a sum of 'fermionic' and 'bosonic' currents $I = I_\mathrm{F} + I_\mathrm{B}$, one obtains using Matsubara Green's functions techniques
\begin{widetext}
\ba
\label{eq:Ix}
  I_\mathrm{F} &&= - 2 \sum_{kl} \left| M_{kl} \right|^2 \mathrm{Im}\left\{ \sum_{x_n^\mathrm{g}} n_\mathrm{F} (x_n^\mathrm{g}) G_\mathrm{e}^\mathrm{ret} (m,x_n^\mathrm{g}-\delta_\mathrm{F}) \underset{z=x_n^\mathrm{g}}{\mathrm{Res}} G_\mathrm{g} (n,z) + n_\mathrm{F} (\epsilon_m^\mathrm{e}) G_\mathrm{g}^\mathrm{adv} (n,\epsilon_m^\mathrm{e}+\delta_\mathrm{F} )\right\} \nonumber \\
  I_\mathrm{B} &&= \frac{2\pi}{a_\mathrm{0}} \sum_{qkl} \left| D_{qkl} \right|^2 \left[ n_\mathrm{F} (\epsilon_k^e) + n_\mathrm{B} (q^2 a_\mathrm{1})\right]  \left[ n_\mathrm{F} (E_l+\delta_\mathrm{B}) - n_\mathrm{F} (E_l)\right] \frac{E_l+\epsilon_l^{g'}}{2E_l} \delta (E_l + \epsilon_k^e - q^2 a_\mathrm{1} + \delta_\mathrm{B} ),
\ea
\end{widetext}
where $x_n^\mathrm{g}$ are the poles of the dressed Green's function 
$G_\mathrm{g} (n,z) = (z-\epsilon_n^\mathrm{g}-\Sigma (n,z))^{-1}$ and the
coefficients $a_\mathrm{0}$ and $a_\mathrm{1}$ are obtained from the
linearisation of the boson propagator 
$D(Q)^{-1} \approx a_\mathrm{0} (i \omega - q^2 a_\mathrm{1}) $. The 
self-energy $\Sigma$ receives an imaginary part from the broadening $\gamma$, 
giving also a nonvanishing imaginary part to the poles $x_n^\mathrm{g}$. 
For $\gamma = 0$, the fermionic current $I_\mathrm{F}$ simplifies to the 
standard Fermi Golden rule form. We have neglected the backward process
in the boson term $I_\mathrm{B}$ as the $\ket{e}$-states are initially unoccupied. 
The detunings are $\delta_\mathrm{F} =  \delta_\mathrm{B} = - \mu +\delta - \Delta E_\mathrm{mf}$, 
where $\Delta E_\mathrm{mf} = E_\mathrm{mf}^e - E_\mathrm{mf}^g$ is the difference in the mean field energies 
of the states $\ket{g}$ and $\ket{e}$.

The current $I$ as a function of the probe field detuning $\delta$ is plotted in Fig.~\ref{fig:cur}.
The chosen Feshbach detuning $\nu_0 = 2.0\,E_\mathrm{F}$ corresponds to the BCS side of
the resonance (in the high-$T_\mathrm{c}$ regime, $T_\mathrm{c} \approx 0.27\,E_\mathrm{F}$).
Due to small number of molecules on the BCS side, we use 
$I = I_\mathrm{F} + I_\mathrm{B} \approx I_\mathrm{F}$.
For comparison, we plot also the current neglecting the pseudogap ($\Delta_{pg} = 0$) in the 
lower figure. The main observations from the results are: 1) {\it Effect of mean field interactions} 
is simply a shift of the resonance peak by $\Delta E_\mathrm{mf}$. This is shown by the symmetric 
Lorentzian at the center (solid line). It corresponds to a high temperature ($\sim T_F$) with 
no preformed pairs. (Comparison: for $\Delta_{pg} = 0$, the simple mean field behaviour persists 
until the critical temperature, the solid line Lorentzian is for $T=T_c$.) 2) 
{\it Effect of pairing and excitation gap} appears as asymmetric form and splitting of the peak. 
Pair formation occurs already above the critical temperature, the dashed line for $T=T_c$ 
represents also $T \gtrsim T_c$. Splitting of the peak is due to opening of an excitation gap. 
The side-peak on the left corresponds to breaking a pair: the gap energy $\Delta$ has to be 
provided, causing a shift from the center. The side-peak on the right originates from finite 
temperature quasiparticles, which are excitations with the minimum energy $\Delta$. Energy is 
{\it gained} when transferring them from $\ket{g}$ to $\ket{e}$, therefore this side-peak appears 
on the opposite side of the center position. The quasiparticle side-peak is weaker and vanishes 
with decreasing temperature, leading to an asymmetric form of the total curve. An important 
feature is that the curve is broadened by the finite lifetime of the pseudogap. 
(Comparison: for $\Delta_{pg} = 0$, pair formation occurs only below $T_c$. Moreover, the 
pseudogap lifetime broadening is absent which leads to sharper features, e.g. the two side-peaks 
are more visible. For $\Delta_{pg} = 0$, the existence of the excitation gap equals superfluidity.) 
3) {\it Superfluidity} is signaled by increase in the asymmetry of the peak (c.f. the dotted 
curve at $T=0.5\,T_c$) as will be discussed in the following. 


\begin{figure}
  \includegraphics[width = 7cm]{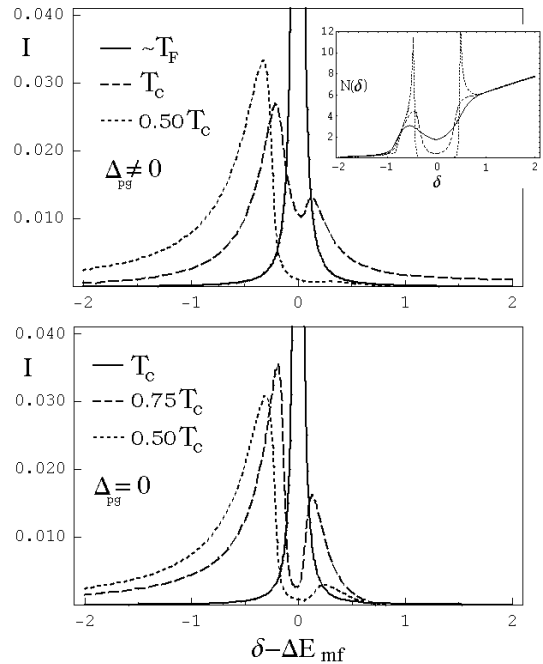}
  \centering
  \caption{The particle current as a function of the probe field detuning $\delta$ for different 
    temperatures. The decay rate of the non-condensed pairs is chosen 
    $\gamma = \Delta (T_\mathrm{c}) \approx 0.6\,E_\mathrm{F}$. For comparison, the 
    lower plot shows
    the current neglecting the pseudogap.  The inset shows the density of states of fermions
    at resonance detuning $\nu_0 = 2.0\,E_\mathrm{F}$. There are two peaks located
    at $\omega = \pm \Delta$, but increasing temperature broadens the peaks. The three
    plots are for temperatures $T = T_c$ (solid), $0.75\,T_\mathrm{c}$ (dashed)
    and $0.10\,T_\mathrm{c}$ (dotted).}
  \label{fig:cur}
\end{figure}

We quantify the asymmetry of the peak $A \equiv \frac{v_\mathrm{left}}{v_\mathrm{right}}$ 
by integrating the current $I$ to the left $v_\mathrm{left}$ and to the right 
$v_\mathrm{right}$ from the peak value. The asymmetry has been plotted as a function of temperature
in Fig.~\ref{fig:asym}, signaling the phase transition by an abrupt change in the asymmetry 
as a function of temperature. The data for temperatures above $T_\mathrm{c}$ have been obtained by
extrapolating the total excitation gap $\Delta (T)$ and solving the chemical potential $\mu$ from 
the number equation. Since the magnitude of the lifetime broadening
$\gamma$ is important for the effectiveness of the method, we show the asymmetry for two
$\gamma$'s of different orders of magnitude; $\gamma \sim \Delta (T=T_\mathrm{c})$ and 
$\gamma \sim 0.2 \Delta (T=T_\mathrm{c})$ in Fig.~\ref{fig:asym}. Time-energy uncertainty relation 
suggests $\gamma = t_{decay}^{-1} \sim \Delta$ which is also used in \cite{Chen2001}.
Note that the change in the asymmetry at $T = T_\mathrm{c}$ is visible also for the
smaller $\gamma$. 
\begin{figure}
  \includegraphics[width = 7cm]{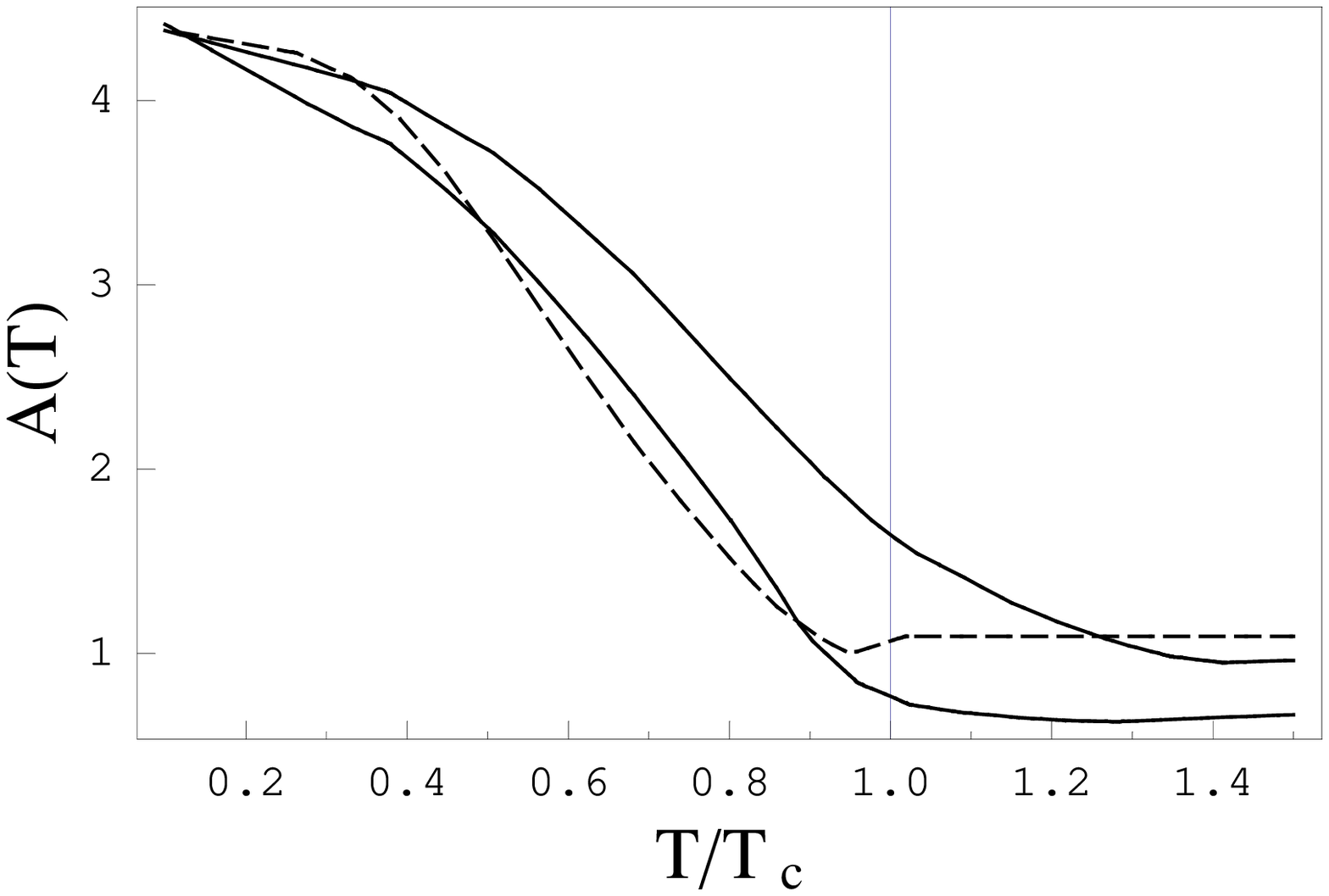}
  \centering
  \caption{The asymmetry $A$ of the current $I$ as a function of temperature 
    $T/T_\mathrm{c}$ for the Feshbach detuning $\nu_\mathrm{0} = 2.0\,E_\mathrm{F}$
    (attractive interaction). Above critical temperature (in the pseudogap regime)
    the asymmetry is essentially constant but rapidly 
    increases in the superfluid regime. The lower solid line is for lifetime broadening 
    $\gamma \approx \Delta (T = T_\mathrm{c})$ and the upper line for 
    $\gamma \approx 0.2 \cdot \Delta (T = T_\mathrm{c})$. For larger $\gamma$,
the pseudogap pairs decay faster and are easier to distinguish from Cooper
pairs, the lower line therefore shows sharper features at $T_\mathrm{C}$. The dashed line neglects the 
    pseudogap, showing the BCS-like behaviour. In the limit 
    $T \rightarrow 0$, the pseudogap vanishes and all three curves approach the same limit. 
    Averaging with relative densities $n=1$, $0.8$ and $0.6$ is included.}
  \label{fig:asym}
\end{figure}

The main experimental challenge related to the method is likely to be mean field effects. 
Figure~\ref{fig:cur} presents results for a homogeneous system where the mean field causes merely 
a shift of the resonance position. In real experiments, the gas is trapped in a harmonic 
potential and consequently the mean field shift varies depending on the local density in 
the trap. This leads to broadening of the shifted resonance peak \cite{rf}. 
In order to estimate trapping effects, in the sense of local density approximation, 
we have calculated the current and the asymmetry for relative densities $n = 1.0$, $0.8$ 
and $0.6$ and mean fields $\Delta E_\mathrm{mf}(n) = 0.5\,E_\mathrm{F}$. Fig.~\ref{fig:asym} 
presents averaged results: the change in the asymmetry is nevertheless clear. The choice 
of the state $\ket{e}$ in such a way that mean field effects are under control, possibly 
combined with schemes that probe only the center of the trap, will be crucial for the success
of the method.

Further challenges appear when going further towards the BCS limit. The obscuring effect of 
the pseudogap and the useful (for this method) effect of the lifetime broadening decrease hand 
in hand, thus in principle there is no essential difference to the scenario presented here 
\cite{footnote2}. However, in the extreme BCS regime only atoms close to the Fermi surface 
are paired and the atoms far from it will be transferred resonantly without the shift of the
peak by the gap energy. In this case, Pauli blocking the lowest energy states for atoms 
$\ket{e}$ by having them initially populated would strengthen the signature.  

In summary, we have studied the possibility of observing resonance superfluidity by a probing field 
transferring atoms out of the superfluid. The emergence of the pseudogap obscures the phase 
transition signature because it makes the transfer resonance peak asymmetric even above $T_c$. 
However, the incoherent nature of the pseudogap broadens the quasiparticle peaks in the spectral 
densities, allowing one to distinguish between the pseudogap and superfluid regimes. 

{\it Acknowledgements} We thank J. Milstein for help with the numerical work. Support is acknowledged from the Academy of Finland (Project No.\ 53903) and European Commission IST-2001-38877 (QUPRODIS).


\begin{thebibliography}{99}
\bibitem{footnote1}
This is small on the BCS side of the resonance due to small number of condensed molecules.
\bibitem{footnote2}
Also on the BEC side ($\nu_0 < 0$) the fermionic current behaves qualitatively as in 
Figs.~\ref{fig:cur} and~\ref{fig:asym}, but the bosonic current $I_\mathrm{B}$ is not
negligible.
\end{thebibliography}
\end{document}